\documentclass{elsart}
\usepackage{epsf}

\def\spose#1{\hbox to 0pt{#1\hss}}
\def\ltapprox{\mathrel{\spose{\lower 3pt\hbox{$\mathchar"218$}}
 \raise 2.0pt\hbox{$\mathchar"13C$}}}
\def\gtapprox{\mathrel{\spose{\lower 3pt\hbox{$\mathchar"218$}}
 \raise 2.0pt\hbox{$\mathchar"13E$}}}
\def\inapprox{\mathrel{\spose{\lower 3pt\hbox{$\mathchar"218$}}
 \raise 2.0pt\hbox{$\mathchar"232$}}}
 \newcommand{\ba}{\begin{eqnarray}}
 \newcommand{\ea}{\end{eqnarray}}
 \newcommand{\be}{\begin{equation}}
 \newcommand{\ee}{\end{equation}}

    \newcommand{\bx}{{\bf x}}

    \newcommand{\matrixele}[3]{\ensuremath{\langle #1 \mid #2 \mid #3\;\rangle}}
\begin{document}
\begin{frontmatter}
\title{A Lattice Study of the Glueball Spectrum$^\dagger$}
\author[pku]{Chuan Liu $^\diamondsuit$} 
\date{}
\address[pku]{Department of Physics,\\
Peking University, Beijing, 100871, P.~R.~China
}
\begin{abstract}    
Glueball spectrum is studied using an improved gluonic action on
 asymmetric lattices in the pure $SU(3)$ gauge theory. The smallest spatial
lattice spacing is about $0.08fm$ which makes the extrapolation to the
continuum limit more reliable. 
In particular, attention is paid to the scalar glueball
mass which is known to have problems in the extrapolation.
Converting our lattice results to physical units using the
scale set by the static quark potential,
 we obtain the following results for the glueball masses:
$M_G(0^{++})=1730(90)MeV$ for
the scalar glueball mass and $M_G(2^{++})=2400(95)MeV$ for
the tensor glueball.\hfill\\[2ex]

\textit{PACS:} 12.38.Gc, 11.15.Ha, 12.39.Mk\hfill\break     
\textit{Keywords:} Glueball spectrum, Lattice QCD.

\end{abstract}
\end{frontmatter}
\vfill
\rule{2.5cm}{0.5mm}\\
{\makebox[5mm][l]{$\diamond$}Email: chuan@mail.phy.pku.edu.cn}\\
{\makebox[5mm][l]{$\dagger$}Work supported by the Chinese
Natural Science Foundation under Grant No. 17500000, Pandeng fund, Startup fund
from MOE and the Startup fund from Peking University.}

\section{Introduction}
\label{sec:intro}

It is believed that QCD is the theory which describes
strong interactions among quarks and gluons. A direct consequence of
this is the existence of excitations of gluonic degrees of freedom,
i.e. glueballs. However, due to their non-perturbative nature,
the spectrum of glueballs can only be investigated reliably with
non-purterbative methods like lattice QCD %
\cite{bali93,michael89,berg83a,berg83b}. Recently, it has become clear
that such a calculation can be performed on a relatively coarse lattice
using an improved gluonic action on asymmetric lattices %
\cite{lepage95:pc,colin97,colin99}. In this paper,
we present our results for a glueball spectrum calculation.
The spatial lattice spacing in our simulations ranges from $0.08fm$ to
$0.25fm$ which enables us to extrapolate more reliably to the
continuum limit.
The improved gluonic action we used is the tadpole improved gluonic
action on asymmetric lattices as described in \cite{colin97,colin99}.
It is given by:
\ba
S=&-&\beta\sum_{i>j}
\left[{5\over 9}{TrP_{ij} \over \xi u^4_s}
-{1\over 36}{TrR_{ij} \over \xi u^6_s}
-{1\over 36}{TrR_{ji} \over \xi u^6_s} \right] \nonumber \\
&-&\beta\sum_{i} \left[
{4\over 9}{\xi TrP_{0i} \over  u^2_s}
-{1\over 36}{\xi TrR_{i0} \over u^4_s} \right] \;\;.
\ea
In the above expression, $\beta$ is related to the bare gauge coupling,
$\xi=a_s/a_t$ 
is the (bare) aspect ratio of the asymmetric lattice with
$a_s$ and $a_t$ being the lattice spacing in spatial
and temporal direction respectively. The parameter $u_s$ is the
tadpole improvement parameter to be determined self-consistently from
the spatial plaquettes in the simulation. $P_{ij}$ and $P_{0i}$ are the 
spatial and temporal plaquette variables. 
$R_{ij}$ designates the $2\times 1$
Wilson loop ($2$ in direction $i$ and $1$ in direction $j$).
Using spatially coarse and temporally fine lattices  helps
to enhance signals in the glueball correlation
functions. Therefore, the bare aspect ratio
is taken to be some value larger than one. In our simulation, we
have used $\xi=3$ for our glueball calculation.
It turns out that, using the non-perturbatively
determined tadpole improvement \cite{lepage93:tadpole}
 parameter $u_s$, the renormalization
effects of the aspect ratio is small \cite{colin97,colin99}, 
typically of the order of a few percent for practical 
values of $\beta$ in the simulation. 
This could also be
verified by measuring corresponding Wilson loops, which will
directly yield the renormalized aspect ratio. For the moment,
we will ignore their difference and simply use the bare value
of $\xi$. It is also noticed that, without tadpole
improvement, this renormalization effect could be as large
as $30$ percent \cite{colin97,colin99}.

This paper is organized as follows. In the next section, we will
explain our calculation of the Wilson loops, glueball correlation
functions which give us the mass values of the glueballs in 
various symmetry sectors. Finite volume errors are discussed and
more importantly, finite lattice spacing errors are analyzed. Special
attention is paid to the scalar sector where the extrapolation 
used to be troublesome at coarse lattices.
In the last section, we will have some discussion of out result
and conclude.

\section{Monte Carlo Simulations}
\label{sec:oneloop}

Glueball spectrum calculations in pure gauge theory basically
involve three stages. The first stage, 
gauge field configurations are generated using some algorithm.
We have utilized a Hybrid Monte Carlo algorithm to update gauge
field configurations. Several lattice sizes have been
simulated and the detailed information can be found in Table.1.
\begin{table}[htb]
\caption{Simulation parameters and the corresponding
lattice spacing in physical units obtained from Wilson loop
measurements.}
\vspace{3mm}
\begin{center}
\begin{tabular}{|c|c||c|c|c|}
\hline 
      Lattices & $\beta$ & $\lambda_W$ & $n_W$ & $r_0/a_s$  \\
\hline 
$ 8^3\times 24$ & $2.4$ & $0.20\sim 0.40$ & $2\sim 4$ & $1.98(2)$ \\
$ 8^3\times 24$ & $2.6$ & $0.20\sim 0.40$ & $2\sim 4$ & $2.48(2)$ \\
$ 8^3\times 24$ & $3.0$ & $0.20\sim 0.40$ & $4\sim 6$ & $4.11(4)$ \\
$ 8^3\times 24$ & $3.2$ & $0.25\sim 0.50$ & $4\sim 6$ & $5.89(8)$ \\
$10^3\times 30$ & $3.2$ & $0.25\sim 0.50$ & $4\sim 6$ & $5.89(8)$ \\
\hline 
\end{tabular}
\end{center}
\end{table}
For each lattice with fixed bare parameters, 
order of $10^3$ configurations have been accumulated. 
Each gauge field configuration
is separated from the previous one by several Hybrid Monte
Carlo trajectories, typically $5\sim 10$, to make sure that
they are sufficiently decorrelated. Further binning of
the data has been performed and no noticeable 
remaining autocorrelation has been observed.

The second stage of the calculation
is to perform measurements of physical observables
using the accumulated  gauge field configurations.
In fact, two independent measurements have to be done. 
One is to set the scale in physical units, i.e.\
to determine the lattice spacing $a_s$ in physical units.
This is necessary since there is no scale in a pure gauge theory.
The second is to measure glueball mass values in lattice
units. This is done by measuring various glueball correlation
functions. With the scale being set in the first step, 
the mass values of the glueballs can then
be converted into physical units. 

In the final stage of the calculation,
glueball mass values obtained from a finite lattice
have to be extrapolated to the continuum limit. This means
that finite volume effects and finite lattice spacing
errors have to be eliminated.
Typically, finite volume errors are found to be small in these
calculations \cite{colin97,colin99}.
It is the finite lattice spacing errors that are more   
difficult to handle, especially for the scalar glueball
mass. It was observed that, the scalar
glueball sector exhibits a dip in the extrapolation, making
the extrapolation less dependable compared with other 
channels \cite{colin97}. To remedy this situation, a simulation at
a smaller values of $a_s$ has been performed. In our calculation,
we also performed a simulation at a smaller lattice spacing where
$a_s \sim 0.08fm$. These two simulations now makes the extrapolation
in the scalar sector more dependable and less sensitive to
the form of the functions used in the extrapolation.

\subsection{Setting the scale}
\label{sec:wilsonloop}
In our simulation, the scale is set by measuring
Wilson loops from which the static quark anti-quark potential
$V(R)$ is obtained. 
Using the static potential between
quarks, we are able to determine the lattice spacing in
physical units  by measuring $r_0$, a pure gluonic scale determined from
the static potential \cite{sommer94:r0}.
The definition of the scale $r_0$ is given
by: $R^2dV(R)/dR]_{R=r_0}=1.65$.
In physical units, $r_0$ is roughly $0.5fm$
which is determined by comparison with potential models. 
For a recent determination of $r_0$, please consult Ref.~\cite{sommer98:r0}. 

In order to measure the Wilson loops accurately, it is 
the standard procedure to perform single link smearing
\cite{michael89,colin97} on the spatial gauge links of the configurations. 
In this procedure, each spatial gauge link is replaced by
a linear combination of the original link and its spatial
staples. Each spatial gauge staple is weighted by a
parameter $\lambda_W$ relative to the original gauge link. The
final result is then projected back into $SU(3)$. 
This smearing scheme can be performed iteratively on
the spatial links of gauge fields for as many as
$n_W$ times. The effect of smearing is to projects out 
higher excitation contaminations
from the Wilson loop measurements. Then, Wilson loops are
constructed using these smeared links. For a Wilson loop
of size $R\times T$, it is fitted against:
\be
W(R,T) \stackrel{T \rightarrow \infty}{\sim} Z(R) e^{-V(R)T}
\;\;.
\ee
Therefore, by extracting the effective mass plateau at large
temporal separation, the static quark potential $V(R)$ is 
obtained. We have measured the Wilson loops along different
lattice axis. It is seen that the measured data points
for the quark anti-quark potential
along different lattice axis lie on a universal
line which is an indication that the improved action restores
the rotational symmetry quite well.
The smearing parameter $\lambda_W$  used in this process are
also listed in Table. 1. Using smearing, we have been able
to obtain descent plateau in the effective mass and the potential $a_tV(R)$ 
is thus determined for various values of $R/a_s$. 

The static quark potential is fitted according to 
a Coulomb term plus a linear confining potential which is known
to work well at these lattice spacings \cite{colin99}. The potential is 
parameterized as:
\be
V(R) = V_0 +e_c/R +\sigma R\;\;.
\ee
From this and the definition of $r_0$, it follows that:
\be
r_0/a_s=\sqrt{{1.65+e_c \over \sigma a^2_s}}
\;\;.
\ee
To convert the measured result to $r_0/a_s$, we have
also used the value of $\xi$ taken as the bare value.
As explained earlier, the renormalization effects for this
parameter is small.
The result of the spatial lattice spacing in
physical units is also included in Table. 1. 
The errors for the ratio $r_0/a_s$ are obtained by blocking 
the whole data set into 
smaller blocks and obtain the error from different blocks.

\subsection{Glueball mass measurement}
\label{sec:correlation}
To obtain glueball mass values, it is
necessary to constructed glueball
operators in various symmetry sectors of interests.
On a lattice, the full rotational symmetry is broken
down to only cubic symmetry, which is a finite
point group with $5$ irreducible representations %
\footnote{Since we only consider the glueball states with
zero momentum, it suffices to study the cubic group. For 
glueball states with non-vanishing momenta, the whole space
group has to be considered.}. They
are labeled as : $A_1$,  $A_2$,  $E$, $T_1$ and $T_2$.
The first two irreducible representations are one-dimensional.
The third is two-dimensional while  the last two are
three-dimensional. In practice, one is interested in the
scalar, tensor and vector glueballs in the continuum limit.
The correspondence is the following: scalar glueball is
in the $A_1$ representation of the cubic group; tensor glueball
is in representation $E+T_2$, which forms a $5$-dimensional
representation; vector glueballs are in the representation $T_1$.
Of course, this correspondence is not one to one but
infinite to one. Therefore, what we can measure is the lowest excitation
in the corresponding representation of the cubic group %
\cite{johnson,berg83a}.

Glueball correlation functions are notoriously difficult to measure.
They die out so quickly and it is very difficult to get a clear
signal. In order to enhance the signal of glueball correlation functions,
smearing and fuzzying have to be performed on spatial links of the
gauge fields \cite{michael89,colin97,colin99}. 
These techniques greatly enhance the overlap of
the glueball states and thus provide possibility of measuring 
the mass values. 
\begin{figure}[htb]
\begin{center}
\epsfysize=5.0cm
\epsfbox{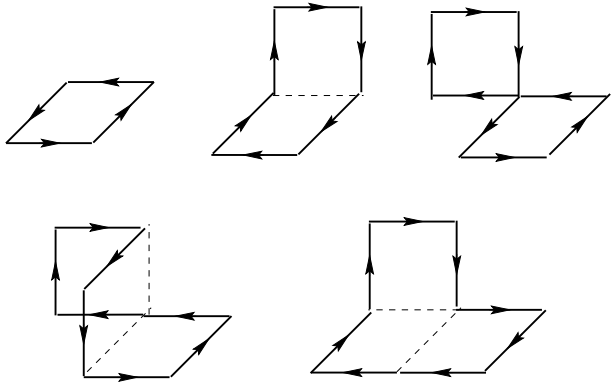}
\caption{The Wilson loop shapes used in constructing the glueball operators.}
\end{center}
\end{figure}
In our simulations, after performing single link
smearing and double link fuzzying on spatial links,
we first construct raw operators at a given time slice $t$:
 $\{ {\mathcal R}^{(i)}_t=\sum_{\bx}{\mathcal W}^{(i)}_{\bx,t}\}$,
where ${\mathcal W}^{(i)}_{t,\bx}$ are closed Wilson loops 
originates from a given
lattice point $x=(t,\bx)$. The loop-shapes studied in this
calculation are shown in Fig.1. Then, elements from the cubic group
is applied to these raw operators and the resulting
set of loops now forms a basis for a representation of the cubic
group. Suitable linear combinations of these operators are
constructed to form a basis for a particular irreducible 
representation of interest \cite{berg83a}.
We denote these operators as $\{ {\mathcal O}^{(R)}_\alpha(t) \}$ where
$R$ labels a specific irreducible representation and $\alpha$ labels
different operators at a given time slice $t$.
In order to maximize the overlap with one glueball state, we construct
a glueball operator
${\mathcal G}^{(R)}(t)=%
\sum_{\alpha}v^{(R)}_\alpha\bar{{\mathcal O}}^{(R)}_\alpha(t)$,
where $\bar{{\mathcal O}}^{(R)}_\alpha(t)=%
{\mathcal O}^{(R)}_\alpha(t)-\matrixele{0}{{\mathcal O}^{(R)}_\alpha(t)}{0}$.
The coefficients $v^{(R)}_\alpha$ are to be determined from a variational
calculation. To do this, we construct the correlation matrix:
\be
{\mathbf C}_{\alpha\beta}(t)=
\sum_{\tau}\matrixele{0}
{\bar{{\mathcal O}}^{(R)}_\alpha(t+\tau)\bar{{\mathcal O}}^{(R)}_\alpha(\tau)}
{0} \;\;.
\ee
The coefficients $v^{(R)}_\alpha$ are chosen such that they minimize the
effective mass 
\be
m_{eff}(t_C)=-{1 \over t_C}\log\left[
{\sum_{\alpha\beta}v^{(R)}_\alpha v^{(R)}_\beta {\mathbf C}_{\alpha\beta}(t_C)
\over
\sum_{\alpha\beta}v^{(R)}_\alpha v^{(R)}_\beta {\mathbf C}_{\alpha\beta}(0) }
\right] \;\;,
\ee
where $t_C$ is time  separation for the optimization. In our simulation
$t_C=1$ is taken. If we denote the optimal values of $v^{(R)}_{\alpha}$
by a column vector ${\mathbf v}^{(R)}$, 
this minimization is equivalent to the following eigenvalue problem:
\be
{\mathbf C}(t_C)\cdot {\mathbf v}^{(R)} 
=e^{-t_Cm_{eff}(t_C)}{\mathbf C}(0)\cdot {\mathbf v}^{(R)} 
\;\;.
\ee
The eigenvector ${\mathbf v}^{(R)}_0$ with the lowest effective
mass then yields the coefficients $v^{(R)}_{0\alpha}$ for
the operator ${\mathcal G}^{(R)}_0(t)$ which best overlaps the lowest
lying glueball in the channel with symmetry $R$.
Higher-mass eigenvectors of this equation will then overlap
predominantly with excited glueball states of a given
symmetry channel.

With these techniques, the glueball mass values are obtained in
\begin{table}[htb]
\caption{Glueball mass estimates for the symmetry channel
$A^{++}_1$, $E^{++}$ and $T^{++}_2$ at various lattice spacings.
The entries corresponding to the highest $beta$ value are the
values after the infinite volume extrapolation. The last row
tabulated the continuum extrapolated result of the glueball mass
values in units of $1/r_0$.}
\vspace{3mm}
\begin{center}
\begin{tabular}{|c|c||c|c|}
\hline 
     $\beta$ &  $a_tM_{A^{++}_1}$ &  $a_tM_{E^{++}}$  &  $a_tM_{T^{++}_2}$ \\
\hline 
 $2.4$ & $0.552(8)$ & $0.980(10)$  & $1.002(9)$ \\
 $2.6$ & $0.482(12)$ & $0.760(16)$  & $0.798(15)$ \\
 $3.0$ & $0.322(8)$ & $0.460(13)$  & $0.470(13)$ \\
 $3.2$ & $0.233(7)$ & $0.323(12)$  & $0.340(10)$ \\
\hline 
 $\infty$ & $4.23(22)$ & $5.77(34)$  & $5.92(32)$ \\
\hline 
\end{tabular}
\end{center}
\end{table}
lattice units and the final results are listed in Table.2. The 
errors are obtained by binning the total data sets into several
blocks and doing jackknife on the blocks. 

\subsection{Extrapolation to the continuum limit}
\label{sec:extrap}
As has been mentioned, finite volume errors are eliminated by
performing simulations at the same lattice spacing but different
physical volumes. This also helps to purge away the possible
toleron states whose energy are sensitive to the size of the volume.
A simulation at a larger volume is done for the smallest lattice
spacing in our calculation. We found that the mass of the scalar
glueball remains unchanged when the size of the volume is increased.
The mass of the tensor glueball  seems to be affected, which is 
consistent with the known result that tensor glueballs have a
rather large size and therefore feel the finiteness of the 
volume more heavily. The infinite volume is obtained by extrapolating
the finite volume results using the relation \cite{luscher86:finitea}:
\be
a_tM^{(R)}(L_s)=a_tM^{(R)}(\infty)
\left(1-\lambda^{(R)}\exp(-\sqrt{3}z/2)/z\right)\;\;,
\ee
where $z=M^{(A^{++}_1)}L_s$. Using the results for the mass of
the $E^{++}$ and $T_2$ glueballs on $8^324$ and $10^330$ lattices
for the same value of $\beta$, the final result for the mass of
these glueball states are obtained.
Glueball mass values for other symmetry sectors are not so sensitive to
the finite volume effects. Therefore, in Table.2, only the extrapolated
values for the smallest physical volume are tabulated. Other entries
are obtained from $8^324$ lattice results.

As for the finite lattice spacing errors, special attention is paid
to the scalar glueball sector where the continuum limit extrapolation 
was known to have problems. Due to the simulation points at small
lattice spacings, around $0.1fm$ and below, the ambiguity in this
extrapolation is greatly reduced. We have tried to extrapolate the
result using different formula suggested in Ref.~\cite{colin99},
the extrapolated results are all consistent within statistical errors.
For definiteness, we take the simple form:
\be
\label{eq:extrap}
r_0M_G(a_s) = r_0M_G(0) + c_1(a_s/r_0)^2 + c_2(a_s/r_0)^4\;\;,
\ee
and
\begin{figure}[htb]
\begin{center}
\epsfysize=10.0cm
\epsfbox{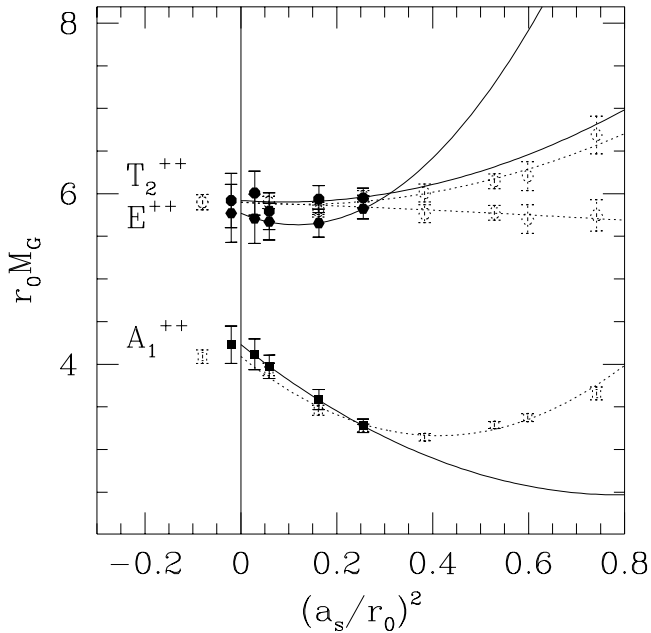}
\caption{The continuum limit extrapolation of glueball mass values
in scalar and tensor channels. The solid symbols are results from
this calculation with the corresponding continuum limit
extrapolation represented by the solid lines. For comparison, the 
corresponding results from \cite{colin99} are also shown with
open symbols and dashed lines.}
\end{center}
\end{figure}
the result is illustrated in Fig.2. The final extrapolated results 
for the glueball mass values are
also listed in Table.2.
The data points from our simulation results are shown
with solid symbols and the corresponding
extrapolations are plotted as solid lines. 
It is also noticed that the extrapolated mass values for
$E^{++}$ and $T^{++}_2$ channels coincide within statistical
errors, indicating that in the continuum limit, they form
the tensor representation of the rotational group.
For comparison, simulation results from \cite{colin99} are
also shown with open symbols and the corresponding
extrapolation are represented by the dashed lines.
We also tried to extrapolate linearly
in $(a_s/r_0)^2$ using three data points with
smallest lattice spacing. The results are statistically consistent
with the results using the extrapolation~(\ref{eq:extrap}) within
errors. It is seen that, due to data points at lattice spacings
around $0.1fm$ and below, the uncertainties in the extrapolation
for the glueball mass values are reduced.

Finally, to convert our simulation results on glueball
masses into physical units, we use the
result $r^{-1}_0=410MeV$. The errors for the hadronic
scale $r_0$ is neglected. For the scalar glueball
we obtain $M_G(0^{++})=1730(90)MeV$. For
the tensor glueball mass in the continuum, we
combine the results for the $T^{++}_2$ and
$E^{++}$ channels and obtain $M_G(2^{++})=2400(95)MeV$ for
the tensor glueball mass.

\section{Discussions and Conclusions}
\label{sec:conclusion}

We have studied the glueball spectrum at zero momentum in
the pure $SU(3)$ gauge theory using
Monte Carlo simulations on asymmetric lattices with the
lattice spacing in the spatial directions ranging from
$0.08fm$ to $0.25fm$. This helps to make extrapolations
to the continuum limit with more confidence
for the scalar and tensor glueball states.
The mass values of the glueballs are converted to physical
units in terms of the hadronic scale $r_0$. 
We obtain the mass for the scalar glueball
and tensor glueball to be:
$m_G(0^{++})=1730(90)MeV$ and $m_G(2^{++})=2400(95)MeV$.
It is interesting to note that, 
around these two mass values, experimental glueball candidates exist.
Of course, in order to compare with the experiments
other issues like the mixing effects and the the effects of quenching have to
be studied.


\end{document}